\documentstyle[mprocl]{article}
\input{psfig}
\def\tg{\widetilde{\gamma}}
\def\tee{\widetilde{\varepsilon}}
\def\ee{\varepsilon}
\def\beq{\begin{equation}}
\def\eeq{\end{equation}}

\def\gsim{\;\rlap{\lower 2.5pt \hbox{$\sim$}}\raise 1.5pt\hbox{$>$}\;}
\def\lsim{\;\rlap{\lower 2.5pt \hbox{$\sim$}}\raise 1.5pt\hbox{$<$}\;}
\def\lesssim{\mathrel{\hbox{\rlap{\hbox{\lower4pt\hbox{$\sim$}}}\hbox{$<$}}}}
\def\gtrsim{\mathrel{\hbox{\rlap{\hbox{\lower4pt\hbox{$\sim$}}}\hbox{$>$}}}}

\begin{document}

\title{Spectral Implications of Variability in GRB 
Fireballs$\,$\footnote{
Talk given at the VIII Marcel Grossmann Meeting on General Relativity,
Jerusalem, June 1997; to appear in the proceedings, eds. R. Ruffini and T.
Piran (World Scientific, Singapore).}}

\author{ Ravi P. Pilla\\
Department of Physics, Columbia University, New York, NY 10027, USA;\\
ravi@cuphyb.phys.columbia.edu\\}

\author{ Abraham Loeb\\
Astronomy Department, Harvard University\\ 60 Garden Street, Cambridge, 
MA 02138, USA;\\
aloeb@cfa.harvard.edu}

\maketitle\abstracts{\hspace*{1cm}
Cosmological $\gamma$-ray bursts originate from relativistic winds.
Temporal fluctuations in the wind velocity can give rise to internal
shocks which dissipate a significant fraction of the wind kinetic energy.
Part of the energy dissipated is transferred to the electrons through Fermi
acceleration. If the post shock fluid is strongly magnetized, the
relativistic electrons cool initially through synchrotron emission, and
later through Compton scattering.  The upsacttered radiation triggers a
cascade of $e^{+}e^{-}$-pairs.  We compute the final spectrum 
for a wide range of parameter values for the
emission region. We show that the spectral diversity observed by BATSE can
be naturally explained by emission from internal shocks, which are
associated with the observed source variability.  }

\parindent=1cm
  
\section{Introduction}

\hspace*{1cm} The origin of $\gamma$-ray bursts (GRBs) is still
unknown.  However, independent of the energy source, the GRB emission
itself must come from a highly relativistic wind.  If most of the wind
energy is carried originally in the form of proton kinetic
energy~\cite{mrtp}, one needs to identify a process which would convert
part of this energy into non-thermal $\gamma$-rays, after the expanding
wind had already become optically thin.  On the basis of the observed
temporal fluctuations in the burst data, it has been suggested that the
burst emission is actually produced through {\it internal shocks} in the
expanding wind~\cite{sp}. In this model the central source is compact
($\sim 10^{6-7}$ cm) and emits a relativistic wind of total energy $\approx
10^{52}$ erg over a time $t_{d}\lesssim 10^{2}$ seconds, with negligible
mass of entrained protons ($\lesssim 10^{28}$ g).  Strong temporal
fluctuations in the luminosity of the source produce many thin layers of
$e^{+}e^{-}\gamma$-plasma, or fireball shells, with a varying specific
energy per unit baryonic mass. These shells acquire different Lorentz factors,
and therefore tend to collide at a larger radius, thus producing (internal)
shock waves. The temporal behavior of the bursts produced by multiple
shell-collisions has been studied recently~\cite{sp}. Here we report some
results on the spectra from such collisions; more details of this
calculation will be given elsewhere~\cite{pl}.

\section{Radiation Mechanisms and Burst Spectra}

\hspace*{1cm}Consider two fireball shells of energy 
$\tee_{i}$ and  rest mass $m_{i}
\,\,(i=1,2)$ starting from the same initial radius.
After an initial acceleration phase, the shells reach Lorentz factors
$\tg_{i}\approx
\tee_{i}/m_{i}c^{2}$, where $c$ is the speed of light in vacuum; 
we take $\alpha\equiv\tg_{2}/\tg_{1}>1$. 
If the second shell is released behind the first one after a time $t_{var}$
(typical variability time) in the observer frame (or the rest frame of 
the central source) the two
shells will collide at an observer-frame radius
$r_{c}=2ct_{var}\tg_{1}^{2}\alpha^{2}/(\alpha^{2}-1) $ and move with a
common Lorentz factor
\beq
\tg=[\tg_{1}\tg_{2}(\tee_{1}+\tee_{2})/(m_{1}\tg_{2}+m_{2}\tg_{1})c^{2}]^{1/2}.
\eeq 
The total energy dissipated in the collision, $\tee_{d}$, defines the
dissipation efficiency
\beq
\xi_{d}\equiv\tee_{d}/(\tee_{1}+\tee_{2})=
1-[\delta_{\ee}(1+\delta_{m})^{2}/(1+\delta_{\ee})(\delta_{\ee}+\delta^{2}_{m})
]^{1/2}, 
\eeq
which depends only on the ratios $\delta_{\ee}=\tee_{2}/\tee_{1}$
and $\delta_{m}=m_{2}/m_{1}$.
\begin{figure}[h]
\centerline{\psfig{figure=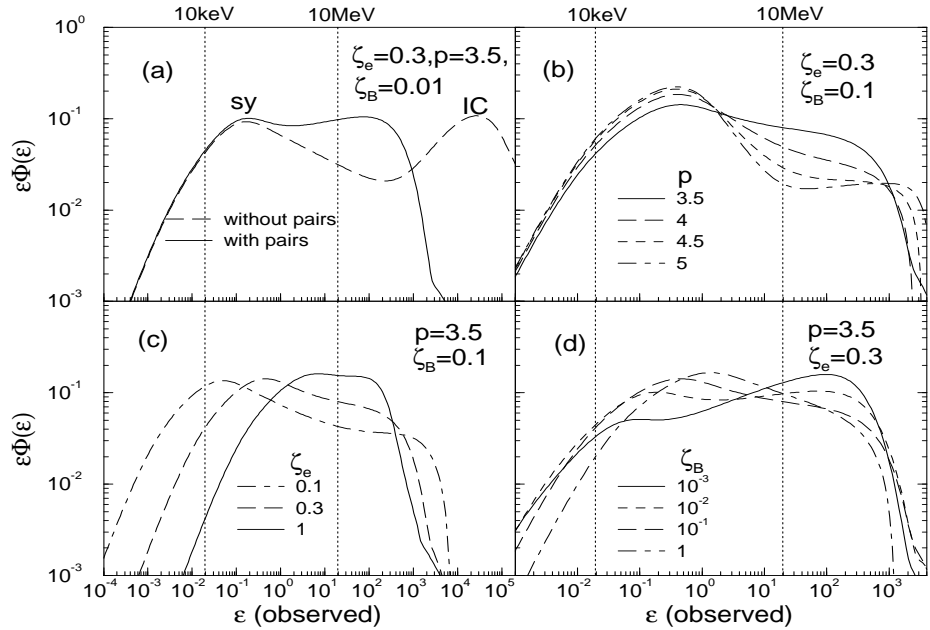,height=3.3in,width=5in}}
\caption{Spectra from internal shocks for physical conditions of the source
given in the text and the conditions in the emitting region specified on
each panel. Panel (a) compares the spectrum before (with synchrotron [sy]
and inverse-Compton [IC] peaks marked) and after the pair cascade.  Panels
(b)-(d) show the dependence of the final spectrum on $p$, $\zeta_{e}$, and
$\zeta_{B}$, respectively.  The results are very sensitive to the value of
$\zeta_{e}$. There is no simple relation between the Fermi index $p$ and
the shape of the spectrum in the BATSE band (marked by the vertical dotted
lines on each
panel).~~~~~~~~~~~~~~~~~~~~~~~~~~~~~~~~~~~~~~~~~~~~~~~~~~~~~~~~~~~~~~~~~~~~~~~~~~~~~~~~~~~~~~~~~~~~~~~~~~~~~~~~~~~~~~~~~~~~~~~~~~~~~~~}
\end{figure}
For $\delta_{\ee,m}\sim$ a few, $\zeta_{d}\gtrsim
10\%$ can be achieved~\cite{sp}. 
The shock waves heat the protons to an average comoving Lorentz factor 
$\overline{\gamma}_{p}\simeq 1+\xi_{d}\tg_{1}(1+\delta_{\ee})/
\tg(1+\delta_{m})$; the protons subsequently  transfer a
fraction $\zeta_{e}$ of their kinetic energy to the electrons through
Fermi acceleration. This transfer of energy is expected to be rapid
[3].

As an example for the typical physical properties of the emission
(post-shock) region, we take a bulk Lorentz factor of $\tg\approx 400$, 
proper density of baryons $n\approx 1.1\times 10^{12}\,\,
\mbox{cm}^{-3}$, comoving thickness $\Delta\approx 4.1\times 10^{10}$ cm,
and $\overline{\gamma}_{p}\approx 3$. With a radiative efficiency of $\sim
100\%$, these parameters correspond to a total observed fluence of $\sim
3\times 10^{-7}\,\,\mbox{erg}/\mbox{cm}^{2}$ for a source redshift
$z_{s}\approx 1$.  Before electron cooling starts, the number of electrons
per unit energy with a Lorentz factor $\gamma$ is assumed to be
proportional to $\gamma^{-p}$, where $p$ is the Fermi-acceleration index;
at that time, the average value of $\gamma$ is
$\overline{\gamma}_{0}\approx 3.6\times 10^{3}\zeta_{e}$. A magnetic
equi-partition fraction $\zeta_{B}$ corresponds to a field strength of
$B\approx 4.7\zeta_{B}^{1/2}10^{4}$ G.  We denote the observer-frame energy
of a photon in units of electron rest mass by $\ee$, and the fraction of
total radiation energy carried by photons in the interval $(\ee,\ee+d\ee)$
by $\Phi(\ee)$, so that $\int_{0}^{\infty}\Phi(\ee)d\ee=1$.  The values of
$p$, $\zeta_{B}$, and $\zeta_{e}$ are free parameters that might vary among
different bursts.

Very high-energy photons are produced as the electrons cool via synchrotron
emission and inverse-Compton scattering on a comoving time scale $t_{c}\sim
1/cn\sigma_{T}\overline{\gamma}^{2}_{0}$, where $\sigma_{T}$ is Thomson
cross section.  Those photons remain in the wind for a time $\sim
t_{0}\equiv \Delta/c\gg t_{c}$, and produce relativistic $e^{+}e^{-}$ pairs
which immediately transfer their energy back to the radiation through
Compton cooling. Such a cascade of pair creation and cooling leaves
pronounced signatures on the emergent spectrum. The final spectra are shown
in Figure 1 for a variety of values for $p$, $\zeta_{B}$, and $\zeta_{e}$.
The additional dependence of the resulting spectra on $r$ and $\tg$ can be
found in [3]. The large variety of possible burst spectra might be
responsible for the spectral diversity of bursts observed by
BATSE~\cite{db}.

\section{Conclusions}

\hspace*{1cm}We have shown that diverse spectra over a wide range of 
energies, covering that of BATSE, is a generic outcome of internal shocks
produced by variability in GRB fireballs. The results shown here only
pertain to the collision of a pair of shells but the qualitative spectral
behavior of the entire event is likely to be similar.

~

~
We thank D. L. Band, E. Cohen, R. Narayan, T. Piran, M. A. Ruderman, 
and R. Sari for useful remarks. 
RP was supported in part by NASA grant NAG5-618 and NSF-MG8
travel award; he acknowledges the hospitality of Racah Institute for Physics, 
The Hebrew University. AL was supported in part by NASA ATP grant NAG5-3085
and the Harvard Milton fund.

\end{document}